\documentclass[twocolumn,showpacs,preprintnumbers,amsmath,amssymb,aps,prl,superscriptaddress]{revtex4}
\usepackage{graphicx}
\usepackage[latin1]{inputenc}
\def\tlse{Laboratoire de Physique Th\'eorique, UMR5152
Universit\'e Paul Sabatier, 31062 Toulouse, France.}
\def\bea{\begin{eqnarray*}}
\def\eea{\end{eqnarray*}}
\def\be{\begin{equation*}}
\def\ee{\end{equation*}}
\def\fig{{\small FIG.}}
\newcommand{\bc}{\begin{center}}
\newcommand{\ec}{\end{center}}
\def\Ham{\mathcal{H}}
\def\nh{n_h}
\def\P{\mathcal{P}}
\def\S{\mathbf{S}}
\def\H{\mathbf{H}}
\newcommand{\moy}[1]{\left\langle{#1}\right\rangle}
\def\ups{\uparrow}
\def\downs{\downarrow}

\begin{document}

\author{G.\ Roux}
\email{roux@irsamc.ups-tlse.fr}
\affiliation{\tlse}
\author{S.\ R.\ White}
\affiliation{Department of Physics and Astronomy, University of
California, Irvine CA 92697, USA.}
\author{S.\ Capponi}
\affiliation{\tlse}
\author{D.\ Poilblanc}
\affiliation{\tlse}

\date{\today}

\title{Zeeman effect in superconducting two-leg ladders: irrational
magnetization plateaus and exceeding the Pauli limit.}


\begin{abstract}
The effect of a parallel magnetic field on superconducting two-leg
ladders is investigated numerically. The magnetization curve displays
an irrational plateau at a magnetization equal to the hole
density. Remarkably, its stability is fundamentally connected to the
existence of a well-known magnetic resonant mode. Once the zero-field
spin gap is suppressed by the field, pairs acquire a finite momentum
characteristic of a Fulde-Ferrell-Larkin-Ovchinnikov phase. In
addition, $S^z=0$ triplet superconducting correlations coexist with
singlet ones above the irrational plateau. This provides a simple
mechanism in which the Pauli limit is exceeded as suggested by recent
experiments.
\end{abstract}
\maketitle

Ladder systems, which consists of two or more strongly coupled chains,
have been widely studied because of their exotic
properties~\cite{Dagotto1996}. In particular, two-leg ladders are Mott
insulators at half filling, with a spin gap, and become
superconducting under doping (and pressure)~\cite{Uehara1996}. This
doped spin-liquid is believed to be a paradigm for the Resonating
Valence Bond mechanism of superconductivity proposed by
Anderson~\cite{Anderson1987}. Such a low-dimensional superconductor
provides an appealing situation to study the role of a parallel
magnetic field (Zeeman effect). In a singlet superconductor, the
critical magnetic field which destroys pairing by splitting up the up
and down spin Fermi surfaces is called Pauli limit. The possibility of
exceeding this theoretical limit and/or stabilizing inhomogeneous
superconductivity predicted by Fulde, Ferrell, Larkin and Ovchinnikov
(FFLO)~\cite{FFLO}, have attracted strong interest in low-dimensional
superconductors~\cite{HighMagnet}. In particular, recent experiments
suggest that Pauli limit is exceeded in the ladder material
Sr$_{14-x}$Ca$_x$Cu$_{24}$O$_{41}$
(SCCO)~\cite{Braithwaite2000,Nakanishi2005}. Furthermore, under
magnetic field, insulating ladders can have plateaus for rational
values of the magnetization, depending on the number of legs and on
the interactions~\cite{Cabra1997}. Doped low-dimensional strongly
correlated systems such as doped spin
chains~\cite{Frahm1999,Cabra2000} have also proved to have interesting
magnetization curves with plateaus at irrational values controlled by
hole doping. Recently, Cabra~\emph{et al.}~\cite{Cabra2002} have
predicted such magnetization plateaus in doped Hubbard ladders by
means of Abelian bosonization and a strong-coupling
expansion. However, numerical evidence supporting such a scenario is
lacking.

In this letter, we investigate doped two-leg ladders with a Zeeman
coupling using the density matrix renormalization group
(DMRG)~\cite{dmrg,RMDdmrg}. We show that, for isotropic couplings
giving rise to a superconducting Luther-Emery
phase~\cite{Hayward1995,White2002}, a non-trivial plateau occurs for
the finite magnetization value predicted in Yamanaka-Oshikawa-Affleck
(YOA) theorem~\cite{Yamanaka1997}. We give a simple physical argument
to explain this behavior which is similar to the formation of the
resonant magnetic mode encountered in the study of spin
dynamics~\cite{Poilblanc2004}. We finally discuss the evolution of the
ground state properties under magnetic field. In particular, we focus
on the persistence of superconductivity for magnetic fields larger
than the Pauli limit. This persistence is associated with the
emergence of Cooper pairs with a finite total momentum, a typical
feature of FFLO phases. We also observed the coexistence of singlet
and triplet pairing.

We describe the doped two-leg ladder using the $t-J$ model with the
same couplings along the legs and between the legs. This model is
believed to account for the unconventional superconductivity in
lightly doped ladders, like the one observed in SCCO. We consider a
static and uniform magnetic field $\H$ which couples to the spin
degree of freedom via a Zeeman term. This corresponds to a magnetic
field oriented parallel to the ladder plane to avoid orbital
effect. Hence, we can write the Hamiltonian \bea \Ham &=&
-t\sum_{\moy{i,j},s} \P_G\left[c^{\dag}_{i,s} c_{j,s} + c^{\dag}_{j,s}
c_{i,s} \right] \P_G\\ &&+J\sum_{\moy{i,j}} [ \S_{i}\cdot\S_{j} -
\frac{1}{4} n_{i} n_{j} ] - g\mu_B \sum_{i} \H \cdot \S_{i} \eea in
which $\P_G$ is the Gutzwiller projector, $c^{\dag}_{i,s}$, $\S_{i}$
and $n_{i}$ are respectively electron creation, spin and density
operators at site $i$ and $s$ is the spin index. In what follows, the
coupling constant $g\mu_B$ is absorbed in the definition of $\H$. We
computed energies at fixed magnetization using the single center site
method recently proposed by one of us which enabled us to have a
discarded weight of order $10^{-7}\!  -\!  10^{-8}$ with 1600 kept
states and a noise level of $10^{-6}$~\cite{White2005}. Energies are
denoted by $E(\nh,S^z)$ with $\nh$ the number of holes and $S^z$ the
total spin along the field. Open boundary conditions are used, $L$
stands for the ladder length and $\delta = \nh / N_{\textrm{sites}}$
is the hole doping ($N_{\textrm{sites}}=2L$). All data were computed
with $J/t=0.5$ for which the model is known to have dominant
superconducting correlations at low
doping~\cite{Hayward1995,White2002}.

\begin{figure}[t]
\includegraphics[width=5.5cm,angle=270,clip]{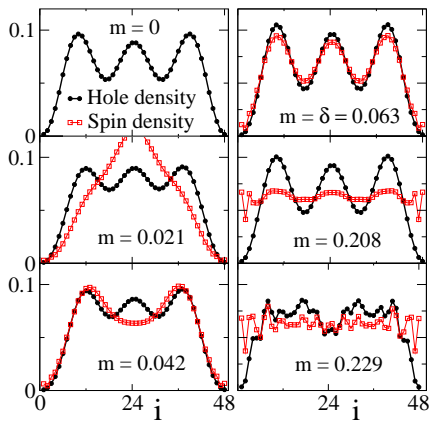}
\caption{(Color online) Evolution of local densities for increasing
magnetization $m$ in a system with 6 holes and $L=48$. The spin
density is normalized by $2\delta/m$ so that spin and hole densities
always have the same mean value.}
\label{densities}
\end{figure}

\begin{figure}[t]
\includegraphics[width=4.5cm,angle=270,clip]{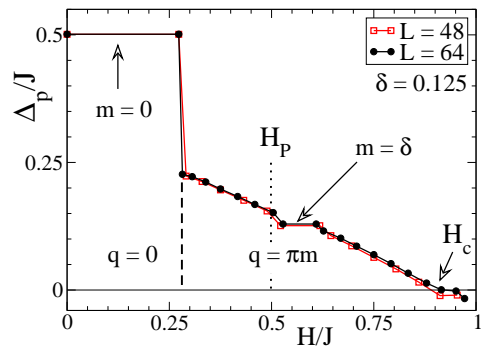}
\caption{(Color online) Pairing energy $\Delta_p$ (see definition in
text) as a function of $\H$ for $\delta = 0.125$. The dashed line
indicates the transition to the FFLO state ($q=0 \rightarrow q=\pi
m$).}
\label{pairing}
\end{figure}

\begin{figure}[t]
\includegraphics[width=4.8cm,angle=270,clip]{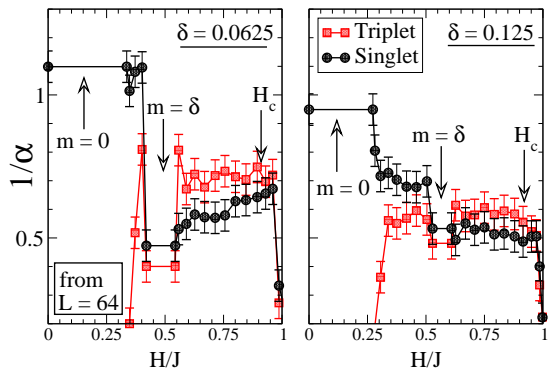}
\caption{(Color online) Inverse of the superconducting correlations
exponent $\alpha$ as a function of $\H$ for two densities. Singlet and
$S^z=0$ triplet correlations coexist above the $m=\delta$ plateau.
Error bars are estimates of uncertainties from fitting.}
\label{exponent}
\end{figure}

\emph{Yamanaka-Oshikawa-Affleck theorem} -- We first recall the
topological constraints that govern the opening of gaps in the
excitation spectrum of one-dimensional like
systems~\cite{Yamanaka1997}. In the case of doped two-leg ladders, YOA
relation takes the simple form \be 1 -\delta \pm m \in
\mathbb{Z}\,,\ee in which $m = 2S^z/N_{\textrm{sites}}$ is the total
magnetization normalized so that it equals $1-\delta$ at
saturation. It is a commensurability condition for either spin $\ups$
or $\downs$ but it introduces irrationality through $\delta$. For
$\delta$ small enough, the relation can be rewritten \be m=\delta\quad
\textrm{or}\quad m=1-\delta\,.\ee The theorem states that: $(i)$ if
the magnetization-doping relation is not satisfied, the low-energy
spectrum can be either a continuum or gapped, but this latest case is
possible only if $1 - \delta \pm m = p/q \in \mathbb{Q}$ and is
associated with a spontaneous breaking of the translational symmetry
so that the ground state is $q$-fold degenerate, $(ii)$ if the
relation is satisfied, a gap \emph{can} open in the excitation
spectrum. Note that for doped systems with charge and spin degrees of
freedom, one sector can be gapped while the other remains
gapless. This actually happens for superconducting two-leg ladders:
the spin gap (related to the $m=0$ plateau) survives for arbitrary
$\delta \neq 0$ but the theorem is still valid because a gapless
charge mode appears simultaneously. This is the so-called Luther-Emery
phase describing the system when $\H=0$.\\
\begin{figure}[t]
\includegraphics[width=5.8cm,angle=270,clip]{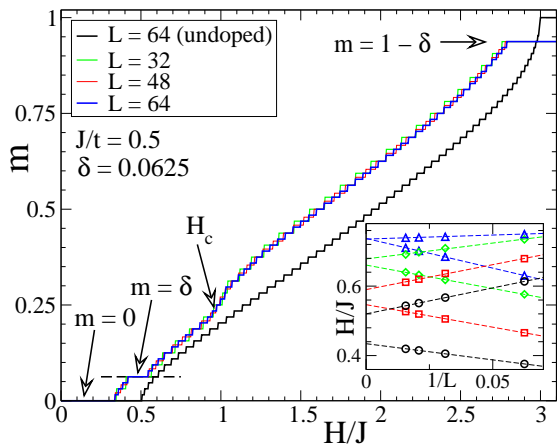}
\caption{(Color online) Magnetization curves at fixed hole density
$\delta=0.0625$ for different system sizes. Three plateaus are visible
for $m=0,\delta$ and $1-\delta$. $H_c$ is the superconducting critical
field (see text) and results in a discontinuity of the slope in the
magnetization curve. The undoped case is shown for
comparison. \emph{Insert :} Finite-size scaling of the two critical
fields corresponding to the boundaries of the $m=\delta$ plateau for
various densities $\delta = 0.0625 (${\Large $\circ$}$), 0.125
(\square), 0.1875 (\Diamond), 0.25 (\triangle)$.}
\label{magnetcurve}
\end{figure}
\begin{figure}[t]
\includegraphics[width=5.8cm,angle=270,clip]{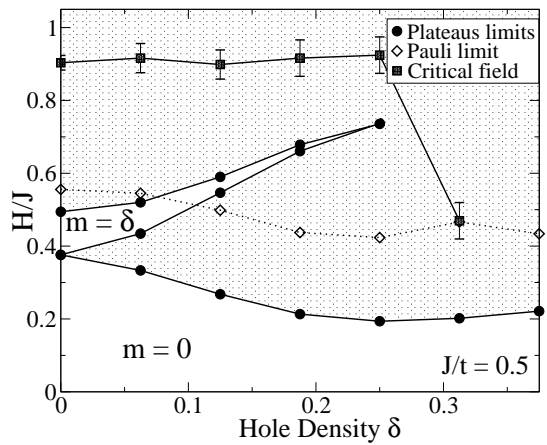}
\caption{Plateau phase diagram in the $(\H,\delta)$ plane. The
expected Pauli limit $H_p$ (see text) is shown in dashed lines. The
superconducting critical field $H_c$ is also shown (the error bars
indicate the uncertainty in determining the transition).}
\label{phasediag}
\end{figure}
\indent\emph{Plateau phase diagram} -- The magnetization curve for a
fixed hole density is given in \fig~\ref{magnetcurve}. Three plateaus
are visible at $m=0,\delta$ and $1-\delta$. The first one corresponds
to the well known spin gap. The second one is expected from the YOA
relation and is the most interesting case. The third one simply
corresponds to the full saturation of the spins. In addition to these
plateaus, the magnetization curve displays a discontinuity in its
slope at a magnetic field $H_c$ above the $m=\delta$ plateau. We will
see that this critical field (consistent with a second order
transition) is the superconducting critical field of the system. The
critical fields of the transitions delimiting the plateaus have been
extrapolated to the thermodynamic limit for various densities (see
\fig~\ref{magnetcurve}) and are gathered in \fig~\ref{phasediag}. The
$\delta\rightarrow 0$ limit has been taken with 2 holes and
$L\rightarrow\infty$. The continuous behavior vs $\delta$ is
consistent with a doping-dependent magnetization
plateaus~\cite{Cabra2000,Cabra2002}. The width of the $m=\delta$
plateau is much smaller than the width of the $m=0$ plateau and
vanishes for $\delta \simeq 0.25$. In the $\delta \rightarrow 0$
limit, it is exactly the binding energy of the hole pair-magnon bound
state previously discussed~\cite{Poilblanc2004}. This bound state
originates from the coupling of the magnon to the charge mode which
leads to an abrupt decrease of the spin gap in the $\delta \rightarrow
0$ limit. Therefore, the formation of the $m=\delta$ plateau is based
on a similar mechanism. To illustrate this, the local hole and spin
densities in the ground state have been computed by DMRG and are
displayed in \fig~\ref{densities} for increasing magnetization on a
system with 6 holes. For the chosen parameters, holes are paired up
and one can see three maxima in the hole density, which qualitatively
correspond to each pair of holes (\fig~\ref{densities}, $m=0$). When a
single magnon is present in the system (\fig~\ref{densities},
$m=0.021$), it preferably binds to a pair of holes. This is due to the
fact that holes gain kinetic energy in a ferromagnetic environment. As
more magnons are added, more bound states progressively form until
$m=\delta$, i.e. when the number of magnons equals the number of hole
pairs. Adding another magnon in this system costs a finite energy gap
leading to the $m=\delta$ plateau. Note that other mechanisms could
lead to plateaus in this system but they would correspond to different
parameters in the Hamiltonian. By analyzing local densities beyond the
$m=\delta$ plateau, one observes that hole pairs survive up to a
magnetization $m \simeq 0.22$. Above the corresponding critical field,
consistent with the value of $H_c$ in \fig~\ref{magnetcurve}, the
system behaves qualitatively like a gas of decoupled spinons and
holons (\fig~\ref{densities}, $m = 0.229$).\\
\indent To confirm the fact that $H_c$ is the superconducting critical
field, we have first computed the pairing energy $\Delta_p$ as a
function of magnetization using \bea \Delta_p(S^z) &=&
E(\nh-1,S^z+1/2) + E(\nh-1,S^z-1/2)\\ && - E(\nh,S^z) - E(\nh -
2,S^z)\,.\eea The results plotted at fixed density on
\fig~\ref{pairing} show that pairing survives far beyond the
plateau. We will see that the superconducting correlations are also
significant up to $H_c$. The $H_c$ line on \fig~\ref{phasediag} was
evaluated from points where $\Delta_p(S^z) = 0$ on a system with
$L=64$. To estimate the Pauli limit in this system, we identify the
condensation energy $\Delta_p^0$ at $\H=0$ to $\Delta_p (S^z =
0)$. Comparing it with the Zeeman stabilization of two up-electrons
gives the Pauli limit $H_P=\Delta_p^0$. Extrapolated results are shown
in \fig~\ref{phasediag}. The fact that the actual superconducting
critical field $H_c$ is larger than $H_P$ suggests that the Pauli
limit is exceeded in this system. We will see that this can be
explained by the appearance of FFLO-like pairing above the $m=0$
plateau.\\
\begin{figure}[t]
\includegraphics[width=5.2cm,angle=270,clip]{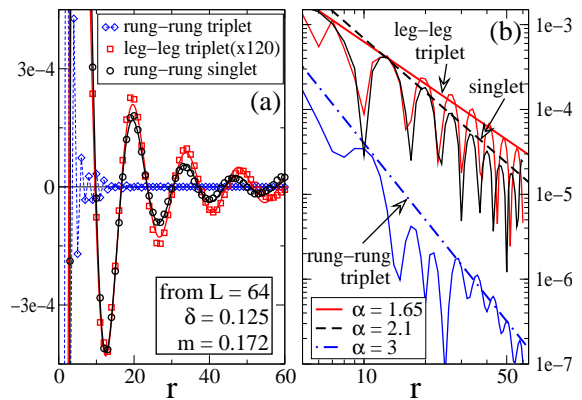}
\caption{(Color online) {\bf(a)} Superconducting correlations in real
space showing $q = \pi m$ oscillations for rung-rung singlet and
leg-leg triplet ($S^z = 0$) correlations while rung-rung triplet ($S^z
= +1$) correlations decay faster (continuous lines are fits). {\bf
(b)} same in log-log scale showing the algebraic decaying with
exponent $\alpha$ (straight lines are envelopes of fits).}
\label{correlations}
\end{figure}
\indent\emph{Evolution of the ground state} -- In order to have a
better characterization of the nature of the ground state under a
magnetic field, we have calculated superconducting correlations
$\langle \Delta_{\sigma\sigma'}^{\lambda} (r)
\Delta_{\sigma\sigma'}^{\lambda\dag} (0) \rangle$ where $\lambda$
characterizes the pairing channel, having a discarded weight of order
$10^{-6}$ with 1000 kept states. Both singlet and triplet channels
have been studied for nearest-neighbor (NN) pairs that can be either
along a rung (called ''rung'' pair) or on the same leg (called ''leg''
pair). When $\H \neq 0$, the four types of spin-pairing are
inequivalent \bea &\textrm{singlet: }& \Delta_{\ups\downs}^s (r) =
c_{r\ups} c_{r+p,\downs} - c_{r\downs} c_{r+p,\ups} \\
&\textrm{triplet: }&\left\{ \begin{array}{l} \Delta_{\ups\downs}^t (r)
= c_{r\ups} c_{r+p\downs} + c_{r\downs} c_{r+p\ups}\\
\Delta_{\ups\ups}^t (r) = c_{r\ups} c_{r+p\ups}\; \textrm{and} \;
\Delta_{\downs\downs}^t (r) = c_{r\downs}c_{r+p\downs}
\end{array}\right. \eea in which $p$ determines the position of the
partner (on same rung or same leg). In the $m=0$ plateau, two-leg
superconducting ladders have anisotropic $d$-wave singlet
pairing~\cite{Hayward1995,Sigrist1994}. Rung-leg correlations and
leg-leg/rung-rung correlations thus have an opposite sign in the
singlet channel. Above the $m = 0$ plateau, we found that both singlet
and leg-leg $S^z=0$ triplet channels give rise to dominant
correlations which long-range behavior is $\cos(qr) / r^{\alpha}$ (see
\fig~\ref{correlations}). The wave vector of the oscillations varies
as $q = \pi m$. This can be qualitatively understood by considering
the Fermi momenta of paired electrons $k_F^{\ups,\downs} =
\frac{\pi}{2} (1 - \delta \pm m)$ on a single
chain~\cite{note2}. Cooper pairs with $S^z=0$ are formed with a total
momentum $q = k_F^{\ups} - k_F^{\downs} =\pi m$ characteristic of FFLO
phases~\cite{FFLO}. We extracted the exponent $\alpha$ to determine
the strength of the correlations (see \fig~\ref{exponent}). In spite
of the difficulty of extracting precisely $\alpha$, we found clear
evidence that both channels are coexisting above the plateau
(although, because of remaining finite size effects, we cannot
conclude on the definite dominance of one w.r.t. the
other~\cite{note2}). Such a rich behavior emerging under a parallel
magnetic field might be surprising at first sight. However, a strong
field dependence of the superconducting correlations is in fact
expected since pairing originates from magnetic fluctuations. In the
irrational plateau, weaker superconducting correlations are found
together with an anomaly in the compressibility (not shown). The
latter could be explained from the finite energy gap needed to remove
a pair of holes at \emph{fixed} magnetization (leading to a deviation
from the $m=\delta$ stability condition). Dominant charge density wave
(CDW) \emph{correlations} could then be expected in this
plateau~\cite{note2}. Note also that the overall magnitude of the
triplet signal is about a hundred times smaller than the singlet one
probably due to a more complex internal pairing structure which NN
pairs only approximate. The leg-leg triplet pairing is symmetric vs
the exchange of chains but further investigations have to be done to
determine the precise orbital structure.\\
\indent\emph{Experiments} -- Experiments in the superconducting phase
of the ladder material SCCO are difficult because of high pressure and
the presence of a chain subsystem. Transport experiments have found
that the superconducting critical field is strongly anisotropic and
that the Pauli limit is likely to be
exceeded~\cite{Braithwaite2000,Nakanishi2005}. We would like to point
out that this behavior does not necessarily imply polarized triplet
pairing \emph{at $\H=0$} or spin-orbit coupling. Indeed, this study
suggests that ladder systems should have a wide FFLO phase explaining
the exceeding of the Pauli limit. Also, possible observation of
irrational magnetization plateaus could determine the hole doping
$\delta$ in ladders. Note that a NMR study proposes $\delta\simeq0.1$
in the superconducting phase~\cite{Piskunov2005}.\\
\indent\emph{Conclusion} -- In conclusion, we show that irrational
magnetization plateaus predicted in~\cite{Cabra2002} do appear in the
study of the isotropic $t-J$ model on two-leg ladders.  Furthermore,
the system has a superconducting critical field larger than the Pauli
limit which can be explained by the emergence of pairing with a total
finite momentum $q=\pi m$ typical of a FFLO phase. This possible
explanation for experiments does not resort to spin-orbit coupling nor
triplet superconductivity at $\H=0$. In addition, under high magnetic
field, we found the emergence of $S^z = 0$ triplet pairing coexisting
with singlet pairing.\\
\indent\emph{Acknowledgments} -- We would like to thank Manfred Sigrist and
Andreas L\"auchli for very fruitful discussions. G.~R. thanks IDRIS
(Orsay, France) for use of supercomputer facilities.
S.~R.~W. acknowledges the support of the NSF under grant DMR03-11843.
We thank ANR for support.


\end{document}